\documentclass[%
 reprint,
superscriptaddress,
showpacs,preprintnumbers,
 amsmath,amssymb,
 aps,
prb,
floatfix,
]{revtex4-1}

\usepackage{graphicx}
\usepackage{dcolumn}
\usepackage{bm}
\usepackage{color}

\usepackage{hyperref}
\hypersetup{
    unicode=false,          
    pdftoolbar=true,        
    pdfmenubar=true,        
    pdffitwindow=false,     
    pdfstartview={FitH},    
    pdftitle={My title},    
    pdfauthor={Author},     
    pdfsubject={Subject},   
    pdfcreator={Creator},   
    pdfproducer={Producer}, 
    pdfkeywords={keyword1} {key2} {key3}, 
    pdfnewwindow=true,      
    colorlinks=true,       
    linkcolor=blue,          
    citecolor=blue,        
    filecolor=blue,      
    urlcolor=blue,       
    plainpages=false        
}

\begin{document}
\title{Magnetic excitations and amplitude fluctuations in insulating cuprates}
\date{\today}
\author{N. Chelwani}
\affiliation{Walther Meissner Institut, Bayerische Akademie der Wissenschaften, 85748 Garching, Germany}
\affiliation{Fakult\"at f\"ur Physik E23, Technische Universit\"at M\"unchen, 85748 Garching, Germany}
\author{A. Baum}
\affiliation{Walther Meissner Institut, Bayerische Akademie der Wissenschaften, 85748 Garching, Germany}
\affiliation{Fakult\"at f\"ur Physik E23, Technische Universit\"at M\"unchen, 85748 Garching, Germany}
\author{T. B\"ohm}
\affiliation{Walther Meissner Institut, Bayerische Akademie der Wissenschaften, 85748 Garching, Germany}
\affiliation{Fakult\"at f\"ur Physik E23, Technische Universit\"at M\"unchen, 85748 Garching, Germany}
\author{M. Opel}
\affiliation{Walther Meissner Institut, Bayerische Akademie der Wissenschaften, 85748 Garching, Germany}
\author{F. Venturini}
\altaffiliation{{Present address: }Z\"urcher Hochschule f\"ur Angewandte Wissenschaften (ZHAW), Technikumstrasse 9, 8401 Winterthur, Switzerland}
\affiliation{Walther Meissner Institut, Bayerische Akademie der Wissenschaften, 85748 Garching, Germany}
\author{{L. Tassini}}
\altaffiliation{{Present address: MBDA, Hagenauer Forst 27, 86529 Schrobenhausen, Germany}}
\affiliation{Walther Meissner Institut, Bayerische Akademie der Wissenschaften, 85748 Garching, Germany}
\author{A. Erb}
\affiliation{Walther Meissner Institut, Bayerische Akademie der Wissenschaften, 85748 Garching, Germany}
\affiliation{Fakult\"at f\"ur Physik E23, Technische Universit\"at M\"unchen, 85748 Garching, Germany}
\author{H. Berger}
\affiliation{Laboratory of Physics of Complex Matter/FSB, Ecole Polytechnique F\'ed\'erale de Lausanne, 1015 Lausanne, Switzerland}
\author{L. Forr\'o}
\affiliation{Laboratory of Physics of Complex Matter/FSB, Ecole Polytechnique F\'ed\'erale de Lausanne, 1015 Lausanne, Switzerland}
\author{R. Hackl}
\affiliation{Walther Meissner Institut, Bayerische Akademie der Wissenschaften, 85748 Garching, Germany}

\begin{abstract}
  We present results from light scattering experiments on three insulating antiferromagnetic cuprates, YBa$_2$Cu$_3$O$_{6.05}$, Bi$_2$Sr$_2$YCu$_2$O$_{8+\delta}$, and La$_2$CuO$_4$ as a function of polarization and excitation energy {using samples of the latest generation. From the raw data we derive symmetry-resolved spectra.} The spectral shape in $B_{1g}$ symmetry is found to be nearly universal and independent of the excitation energy. The spectra agree quantitatively with predictions by field theory [\onlinecite{Weidinger:2015}] facilitating the precise extraction of the Heisenberg coupling $J$. {In addition, the asymmetric line shape on the high-energy side is found to be related to amplitude fluctuations of the magnetization. In La$_2$CuO$_4$ alone minor contributions from resonance effects may be identified.} The spectra in the other symmetries are not universal. The variations may be traced back to weak resonance effects and extrinsic contributions. For all three compounds we find support for the existence of chiral excitations appearing as a continuum in $A_{2g}$ symmetry having an onset slightly below $3J$. In La$_2$CuO$_4$ an additional isolated excitation appears on top of the $A_{2g}$ continuum.
\end{abstract}
\maketitle


\section{Introduction}
If a continuous symmetry is spontaneously broken the order parameter can fluctuate about either its average direction or magnitude. The changes in direction are usually mass-less (Nambu-Goldstone mode) and decoupled from the amplitude fluctuations (Higgs mode) having a finite energy. The Higgs fluctuations play a crucial role in a variety of interacting systems \cite{Podolsky:2011,Pekker:2015} including the standard model in high-energy physics \cite{Higgs:1964}, Bose condensates \cite{Endres:2012}, super-fluidity and -conductivity \cite{Sooryakumar:1980,Littlewood:1981,Cea:2014} and magnetism \cite{Ruegg:2008}. The discussion picked up new momentum after the discovery of the cuprates and the iron-based compounds for the putative interrelation of magnetic and other types of fluctuations and superconductivity \cite{Perali:1996,Lee:2006,Toyota:2007,Anderson:2007,Mazin:2008,Pfleiderer:2009,Hirschfeld:2011,Scalapino:2012,Fradkin:2015,Pekker:2015,Lederer:2015} since, at least in the cuprates, magnetic fluctuations can be observed in large regions if not in the entire phase diagram \cite{Wakimoto:2007}.

In magnetic systems, the order in the ground state, the exchange energies $J_i$, the excitation spectra or fluctuations of the phase or the amplitude of the magnetization \cite{Zwerger:2004,Podolsky:2011} are still a matter of debate both from an experimental and a theoretical point of view. While neutron and Raman scattering experiments on insulating antiferromagnetic cuprates \cite{Lyons:1988,Sugai:1988,Sulewski:1990} can be described qualitatively in terms of spin-wave theory in a Heisenberg model \cite{Fleury:1968,Shastry:1990,Knoll:1990,Canali:1992,Chubukov:1995b,Rubhausen:1997} with only nearest neighbor exchange coupling $J_1$, the role of multi-magnon and cyclic spin excitations or amplitude fluctuations are not well understood yet \cite{Shastry:1990,Sulewski:1991,Vernay:2007,Podolsky:2011}.

Recently, a field theoretical approach successfully described the line shape of the two-magnon Raman spectra in $B_{1g}$  symmetry considering perturbation theory up to infinite order and amplitude fluctuations of the order parameter. For La$_2$CuO$_4$ (La214) and YBa$_2$Cu$_3$O$_{6.05}$ (Y123) the line shapes can be well reproduced using slightly different masses for the Higgs modes \cite{Weidinger:2015}. Similar agreement was found for iridates \cite{Gretarsson:2016} suggesting the response in $B_{1g}$ symmetry to be universal. Yet resonance effects and extrinsic contributions such as luminescence were found to have a substantial influence on the line shape {\cite{Lyons:1988a,Lyons:1989,Sugai:1990,Yoshida:1992,Blumberg:1996,Muschler:2010a} and need to be considered upon properly interpreting the response}.

{Resonance effects of the Raman response of the simultaneous excitation of two magnons at $\Omega_{\bf q}$ and $\Omega_{-\bf q}$, where $\Omega_{\bf q}$ is the magnon dispersion, were studied theoretically by several authors. Shastry and Shraiman \cite{Shastry:1990} used the one-band Hubbard model as a starting point and described the usual resonances when either the energy of the incident or scattered photon, $\omega_{I,S}$, corresponds to a real transition between the lower and the upper Hubbard band. Chubukov and coworkers \cite{Chubukov:1995a,Chubukov:1995b,Morr:1997} realized that there exists a triple resonance for the parameters of the cuprates when $\omega_{I}$, $\omega_{S}$, and $\omega_I-\Omega_{\bf q}$ simultaneously match transitions between the {valence and the conduction} band. In this case the two-magnon maximum in the Raman spectrum slightly below  $3\,J $ has a resonance energy different from $2\max(\Omega_{\bf q})$ at $4J$. As a consequence the line shape depends on $\omega_I$. Brenig \textit{et al.} \cite{Brenig:1997} derived the cross section from the more realistic three-band Hubbard model. {Zhang and Rice \cite{Zhang:1988} have shown that the two models are equivalent to some extent if the large Hubbard $U$ of 5-10\,eV, being relevant for the cuprates, is replaced by an effective energy $\Delta E$ close to the charge transfer energy between copper and oxygen $|\varepsilon_d-\varepsilon_p|$ in the range of 2\,eV which corresponds to the optical gap.}}

In this publication we present results of light scattering experiments on three different cuprate families, La214, Y123 and Bi$_2$Sr$_2$YCu$_2$O$_{8+\delta}$ (Bi2212:Y). {In addition to the so far published symmetry-resolved response for Gd214 and La214 \cite{Sulewski:1991,Muschler:2010a} we derive and compare symmetry-resolved spectra for the three compound classes La214, Y123, and Bi2212:Y and} use various laser energies {in the range 2.16 and 2.71\,eV} for excitation so as to separate the intrinsic non-resonant line shape from a putative influence of resonant light scattering {and extrinsic contributions}. Below three times the magnetic exchange energy $J$ the spectra are essentially universal in that the response $\chi^{\prime\prime}(\Omega)$ depends only on $J$. Above the two-magnon maximum there may be non-universal contributions which are explained in terms of a Higgs mode and minor contributions from resonant light scattering and luminescence.

\section{Experiment}
Y123 was prepared in BaZrO$_3$ crucibles \cite{Erb:1995} demonstrated to yield the highest purity \cite{Erb:1996}. To obtain the best crystal quality and the lowest possible oxygen content of approximately 6.05 the as-grown material was annealed in Ar at 900\,C {for one day in a tube furnace with the single crystal buried in sintered Y123} and then slowly cooled to room temperature {during another day}. Bi2212:Y was grown in Y-stabilized ZrO$_2$ crucibles \cite{Kendziora:1993} and La214 was prepared using the traveling solvent floating zone (TSFZ) method. The as-grown crystals had a N\'eel temperature $T_N$ of 280\,K. After annealing the crystal in Ar, $T_N$ increased to 325\,K [\onlinecite{Muschler:2010a}] which has not been exceeded so far \cite{Kastner:1998,Gozar:2005}. Given these facts  it is likely that no better samples of Y123 and La214 can be obtained to date. There are only very few reports on the successful preparation of Bi2212:Y, and we do not have data for comparison but the results shown below indicate that these crystals are in the same quality class as Y123 and La214.

For the measurements on Y123 and Bi2212:Y we used as-grown and freshly cleaved surfaces, respectively. The surface for the experiments on La214 was polished so as to obtain sufficiently large areas of approximately $2\times 2\,{\rm mm}^2$. In order to test the surface quality after polishing we cleaved a small piece. The results obtained from the cleaved surface and the polished surface are close to identical. In particular all features appear in the same positions as shown in Fig.~\ref{sfig:cleave} in Appendix \ref{Asec:surface}.

The samples were mounted on the cold finger of a He-flow cryostat in a cryogenically pumped vacuum. The experiments were performed with a calibrated light scattering setup \cite{Opel:2000d,Muschler:2010a} with the resolution set at 20\,cm$^{-1}$ at 458\,nm. For excitation an Ar ion laser (Coherent, Innova 304C) was used emitting at 528, 514.5, {496.5, 488, 476 or 458\,nm. For one experiment on La214 we used a solid state laser (Coherent, Genesis MXSLM) emitting at 575\,nm.} The laser-induced heating was determined experimentally to be close to 1\,K per mW absorbed power. Spectra were measured in the six polarization configurations $xx$, $xy$, $x^\prime x^\prime$, $x^\prime y^\prime$, $RR$, and $RL$, where $x^\prime = 1/\sqrt{2}(x+y)$, $y^\prime = 1/\sqrt{2}(y-x)$, and $R,L = 1/\sqrt{2}(x\pm iy)$, allowing one to extract all in-plane symmetry components, $\mu = A_{1g}$, $A_{2g}$, $B_{1g}$, and  $B_{2g}$, of the tetragonal $D_{4h}$ space group by linear combinations of the experimental spectra. For the polarization and symmetry assignment we use the idealized quadratic CuO$_2$ plane. The spectra we show below represent the response $R\chi^{\prime\prime}(\Omega,T)$ as a function of the Raman shift $\Omega$ and the temperature $T$ that is obtained by dividing the  cross section by the Bose-Einstein factor $\{1+n(T,\Omega)\}=[1-\exp(-\hbar\Omega/k_BT)]^{-1}$. $R$ is an experimental constant.

\begin{figure}[tbp]
  \centering
  \includegraphics[width=0.9\columnwidth]{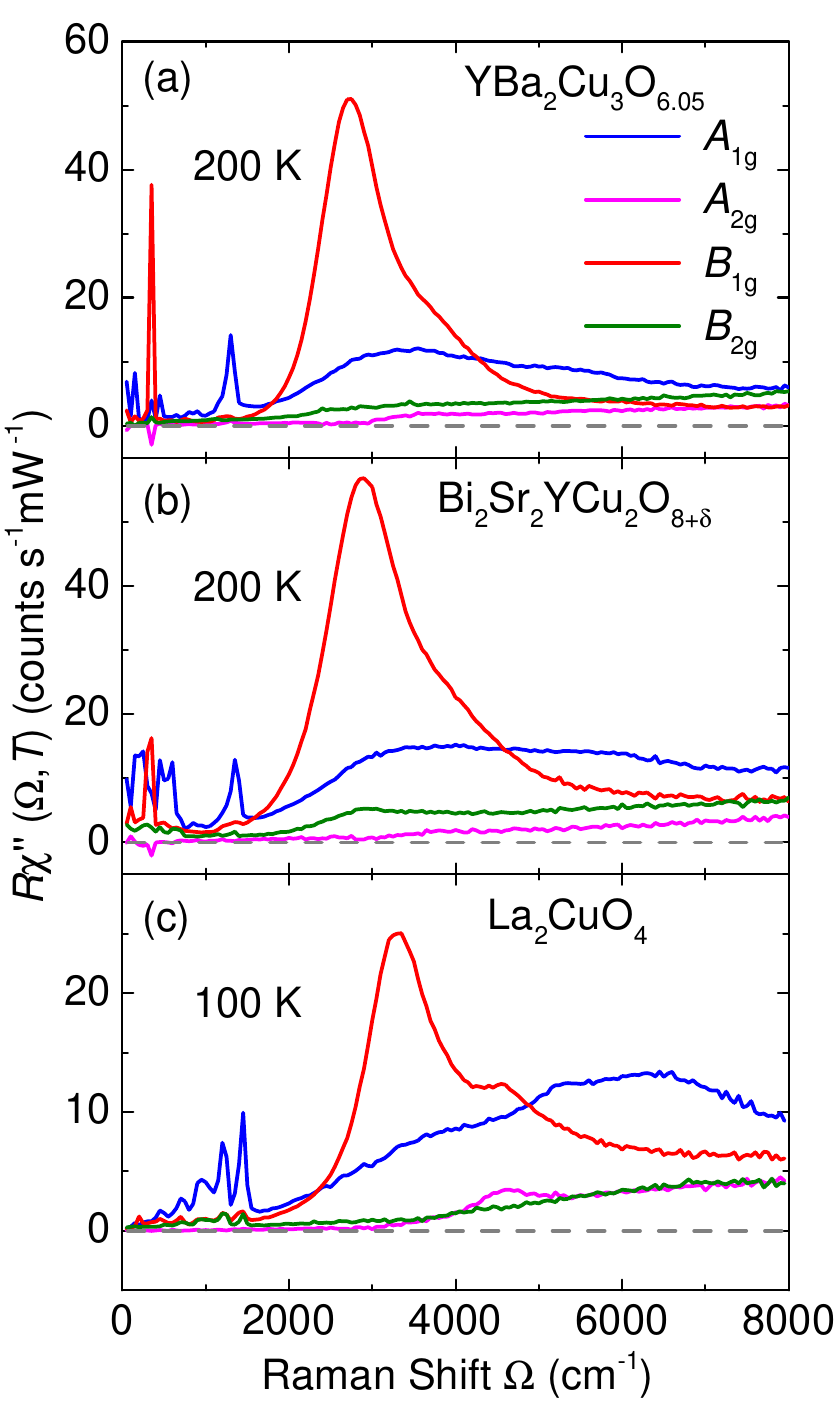}
  \caption{Symmetry resolved Raman response $R\chi_\mu^{\prime\prime}(\Omega,T)$ of ${\rm CuO_2}$ compounds as indicated. The spectra are the result of linear combinations of in total six spectra as described in Appendix \ref{Asec:raw}.
  }
  \label{fig:sym_raw}
\end{figure}

\begin{figure}[tbp]
  \centering
  \includegraphics[width=0.9\columnwidth]{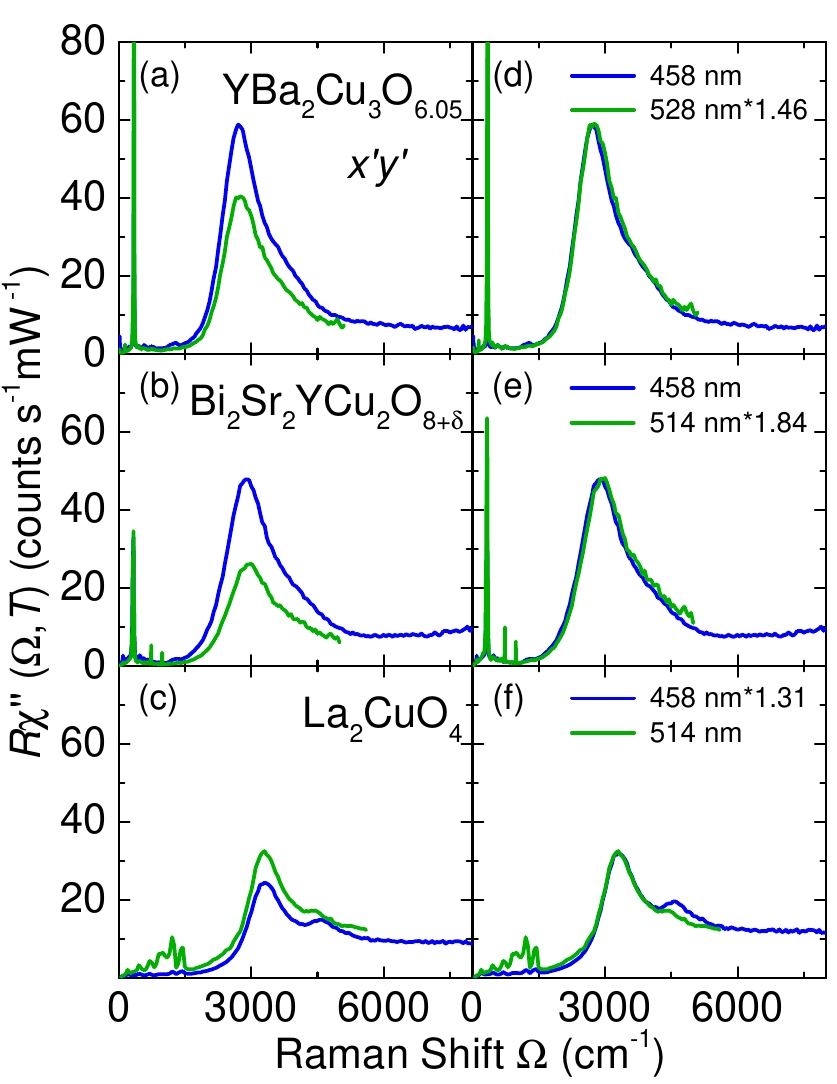}
  \caption{Raman response $R\chi^{\prime\prime}(\Omega,T)$ in $x^\prime y^\prime$ polarization ($B_{1g}+A_{2g}$ symmetry) as a function of excitation energy for Y123, Bi2212:Y and La214. (a)-(c) Raw data. (d)-(f) The spectra are multiplied by factors as indicated so as to match the peak intensities.
  }
  \label{fig:B1g_res}
\end{figure}

\section{Results}
The raw data of La214 measured at all six main polarization combinations are displayed in Fig.~\ref{sfig:raw} in Appendix \ref{Asec:raw}. In Fig.~\ref{fig:sym_raw} we show the spectra of Y123, Bi2212:Y and La214 for all four pure symmetries Raman-active for light polarizations in the CuO$_2$ plane.

The spectra in $B_{1g}$ symmetry have the highest intensity in the range 2000 to 4000\,cm$^{-1}$. The peak maxima of Y123, Bi2212:Y, and La214 are observed to be at 2735, 2890, and 3300\,cm$^{-1}$ (339.3, 358.6, and 409.4\,meV), respectively. At low energies the continuum  is superposed by narrow lines from phonons which are not of interest here and {were} therefore measured with an energy-dependent resolution in the range $10-20$\,cm$^{-1}$ and a step width of 50\,cm$^{-1}$. Below the maximum the spectra depend strongly on energy. The variation proportional to $\Omega^3$ predicted theoretically for low energies \cite{Podolsky:2011,Weidinger:2015} cannot directly be observed because of the contributions from phonons and particle-hole excitations which cannot reliably be subtracted. Above the maximum the intensity decays approximately as expected for a Lorentzian oscillator except for the range around 4000\,cm$^{-1}$ making the peak rather asymmetric. At high energies the intensity becomes very small. Whereas Y123 and Bi2212:Y are similar it is clear at first glance that La214 is different in that a secondary maximum appears above the main peak.

Differences between La214 and the other compounds are also present in the other symmetries. In all materials there is a broad maximum between 3000 and 4000\,cm$^{-1}$ in $A_{1g}$ symmetry, slightly above the position of the $B_{1g}$ peak. In La214 alone there is another broad feature having two substructures between 5000 and 7000\,cm$^{-1}$ which is stronger than the first one. In Bi2212:Y and Y123 there are {just weak} shoulders in this energy range. In Y123 this shoulder is barely visible. No qualitative differences between the compounds are found in $B_{2g}$ symmetry. In the range of the $B_{1g}$ peak the {$B_{2g}$} spectra differ only quantitatively.

In $A_{2g}$ symmetry there are two remarkable observations: (i) In La214 there is a well resolved peak at 4600\,cm$^{-1}$, slightly above that in $B_{1g}$ symmetry, which is completely absent in the other compounds. (ii) {In all compounds} the intensity vanishes to within the experimental resolution for energies {below approximately 3000\,cm$^{-1}$ close to} the $B_{1g}$ maximum. The onset above this energy is well defined and nearly abrupt being reminiscent of gap-like behavior for low energies. {As studied in detail for La214 [Fig.~\ref{sfig:resonance-sym}\,(b)], the onset is independent of the excitation energy indicating an intrinsic scattering process.}

Since the high-energy parts of the $B_{1g}$ spectra are not universal both close to the maximum and at high energies, the origin of the differences needs to be explored before the line shape will be analyzed. To this end excitation with various laser lines is useful. In Fig.~\ref{fig:B1g_res}\,(a)-(c) we show the raw data of the $x^\prime y^\prime$ spectra ($B_{1g}+A_{2g}$) for all the compounds using green and blue excitation. The spectra taken with green excitation end already slightly above 5000\,cm$^{-1}$ Raman shift because the range of the spectrometer is limited to {725}\,nm. In Y123 and Bi2212:Y the scattering cross sections are slightly stronger with blue excitation. In La214 the spectra measured with green light are more intense, and the phonons in the range up to 1500\,cm$^{-1}$ resonate strongly.

For better comparison constant multiplication factors are applied so as to match the peak intensities [Fig.~\ref{fig:B1g_res}\,(d)-(f)]. Obviously the line shape is identical to within the experimental accuracy (statistical error and calibration) for Y123 and Bi2212:Y. For La214 the small extra peak at 4600\,cm$^{-1}$ is weaker with green excitation. As a consequence it moves closer to the main peak. {For better insight we looked at this variation using six different lines as shown in Fig.~\ref{sfig:resonance-x'y'} of Appendix \ref{Asec:resonance-x'y'}. The relative intensity of the extra peak at 4600\,cm$^{-1}$ decreases rather rapidly upon changing the laser line from 488 to 496\,nm [Fig.~\ref{sfig:resonance-x'y'}\,(b)].}

\section{Discussion}

Qualitatively speaking the peak in $B_{1g}$ symmetry comes from a double spin-flip on two neighboring Cu atoms \cite{Fleury:1968,Sulewski:1990}. However, the resulting line shape is a matter of debate as outlined in the introduction. Here we address the question as to whether the line shape is universal or has universal components. There are obvious differences between La214 and the two other materials. To be more quantitative we plot the $B_{1g}$ spectra of Fig.~\ref{fig:sym_raw} on a normalized energy scale.

\begin{figure}[h]
  \centering
  \includegraphics[width=0.95\columnwidth]{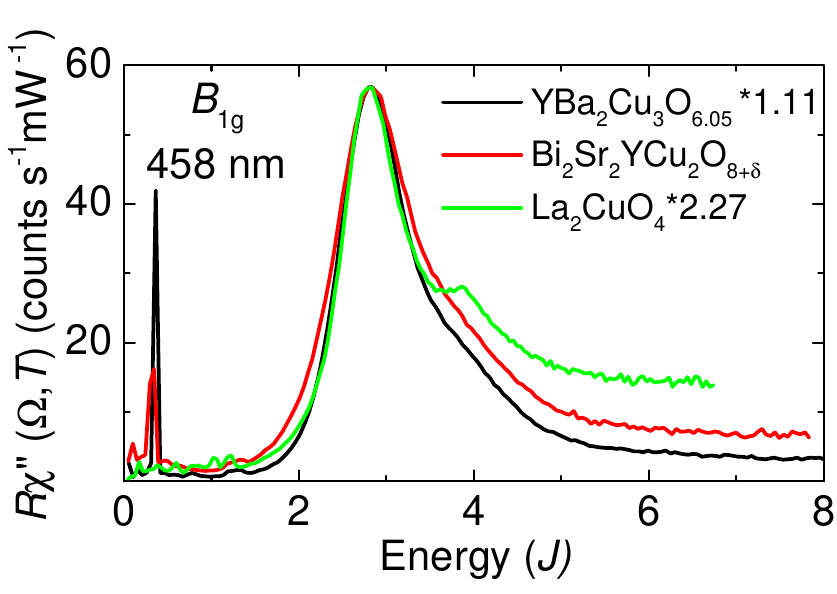}
  \caption{$B_{1g}$ spectra of YBa$_2$Cu$_3$O$_{6.05}$ (black), La$_2$CuO$_4$ (green), and Bi$_2$Sr$_2$YCu$_2$O$_{8+\delta}$ (red). The spectra are reproduced from Fig.~\ref{fig:sym_raw} with the intensities scaled as indicated. The energy axis is given in units of the exchange coupling $J$. The normalized position of the peak maxima matches the theoretically predicted energy of 2.84\,$J$ [\onlinecite{Weidinger:2015}].
  }
  \label{fig:scaling}
\end{figure}

Fig.~\ref{fig:scaling} shows a superposition of the spectra of Y123, La214, and Bi2212:Y. Both the intensity and the energy scales are normalized. The spectra of Y123 and La214 are multiplied by factors of 1.11 and 2.27, respectively, to match the peak intensity of Bi2212:Y. The energies are divided by appropriate factors to collapse the peak positions on the canonical value of 2.84\,$J$ derived recently (There is a typo in the paper by Weidinger and Zwerger.) \cite{Weidinger:2015}. From the peak positions alone one obtains 119, 144, and 126\,meV, respectively, for $J$. For Y123 and La214 the full theoretical analysis including the Higgs mode yields 126 and 149\,meV fairly close to the values of 120 and 143\,meV found by neutron scattering \cite{Reznik:1996,Headings:2010}.

Close to the maximum between 2.5 and 3.2\,$J$  the line shapes of the peaks are rather similar. At low energies the lines of Y123 and La214 coincide down to 2.3\,$J$. Below 2.3\,$J$ La214 has a higher intensity from resonantly enhanced phonon contributions. Above 3.2\,$J$ La214 exhibits a secondary maximum whereas Y123 has only an almost linear tail before the intensity saturates at a value lower than that in all other materials. The two-magnon maximum of Bi2212:Y is generally wider than those of La214 and Y123. The substantially off-stoichiometric composition of Bi2212:Y suggests that the line is inhomogeneously broadened, i.e. the larger width is extrinsic. The most dramatic differences are found above 4.4\,$J$ where the saturation values of the intensities differ by more than a factor of three. We do  not believe that this part of the spectra results from differences in the spin excitations. Rather, contributions from luminescence should be considered a possible (extrinsic) origin.

Concerning the overall intrinsic line shape a satisfactory description was achieved only recently using field-theoretic methods \cite{Weidinger:2015} after various studies on the basis of spin-wave theory \cite{Knoll:1990,Canali:1992,Rubhausen:1996,Rubhausen:1997}. The progress became possible through the inclusion of multi-magnon processes and amplitude (``Higgs'') fluctuations of the staggered magnetization on equal footing up to infinite order as suggested first by Podolsky and coworkers \cite{Podolsky:2011}. As a result the variation of the low-energy response, the nearly Lorentzian variation above the peak and the shoulder between 3.2 and 4.4\,$J$ in Y123 and Bi2212:Y found an almost quantitative explanation. In addition, the differences to La214 can be traced back to a variation of the coupling constant {(``Higgs mass'')}of the {anplitude} mode. The detailed comparison between theory and experiment is shown in Ref.~[\onlinecite{Weidinger:2015}].

The line shape can be explained very well for the La214 data taken with blue excitation {by including amplitude (``Higgs'') fluctuations of the magentization} (see Ref. \onlinecite{Weidinger:2015}). With green excitation the secondary maximum is weaker {and completely absent for red photons \cite{Gozar:2005}}. The dependence of the high-energy peak on the excitation line is not easily explained in terms of amplitude fluctuations since the spins and the amplitude fluctuations are excited via the same intermediate electronic states. Alternatively, the secondary maximum could originate in other types of spin excitations, for instance chiral or multi-magnon excitations that could couple to high-energy electronic states different from those coupling to the double spin-flip excitations. The intensity in $A_{2g}$ symmetry which has a well defined peak in La214 {at a (nearly) constant Raman shift displaying the same resonance behavior as the two-magnon excitation in $B_{1g}$ symmetry (see Appendix \ref{Asec:resonance-sym})} argues for the existence of chiral excitations in addition to double spin flip scattering {rather than an independent type of scattering}. However, the interrelation of the $B_{1g}$ and $A_{2g}$ spectra is not yet clear. First, the similarity of the excitation energies close to {4600\,cm$^{-1}\sim 4\,J$} in both symmetries {may be accidental but} remains to be explained. Second, {since} intensity {from chiral excitations according to the operator $\hat{C}={\bf S}_i\cdot({\bf S}_j\times{\bf S}_k)$} should not appear in first order \cite{Ko:2010,Michaud:2011} {the cross section should be smaller than in all other symmetries}.

No effects similar to those in La214 are observed in Y123 and Bi2212:Y, where the line shapes were found to be independent to within the experimental accuracy {for excitation in the blue and green spectral range. For excitation outside and probably also inside this range variations were found earlier \cite{Blumberg:1996} but it is difficult to compare those data to our results on a quantitative basis.}  Hence, a final experimental answer can still not be given {and requires more work on samples of the latest generation.}

{The} theoretical stud{ies} proposed by Chubukov, Frenkel, and Morr \cite{Chubukov:1995a,Chubukov:1995b,Morr:1997} {qualitatively explain changes of the $B_{1g}$ line shape but make also predictions for the other symmetries which we cannot confirm here: As opposed to the prediction the experimental $A_{1g}$ and $B_{2g}$ spectra {of LCO} resonate towards the green-yellow spectral range as shown in detail in Fig.~\ref{sfig:resonance-sym} whereas the two-magnon peak is strongest for 496\,nm}.

Line shapes of the two-magnon spectra similar to those found for the cuprates in $B_{1g}$ symmetry were also observed in equivalent symmetry projections in some iridates \cite{Gretarsson:2016} suggesting a universal behavior. However, the spectra in the other symmetries and in metallic materials such as doped cuprates or  Fe-based compounds remain widely unexplained and cannot be treated in the same fashion as the $B_{1g}$ spectra of insulating CuO$_2$ using field theory where convergence is granted by the $B_{1g}$ form factor $\propto k_x^2-k_y^2$. For addressing subjects {such} as resonances and the symmetry dependence, only numerical methods are currently available \cite{Moritz:2011,Chen:2011b,Jia:2014}.

\section{Conclusions}
We have measured high-energy spectra of three different cuprates {at all main polarization configurations} {using at least two different excitation lines and, in addition to earlier work, derived all pure} symmetries{, $A_{1g}$, $A_{2g}$, $B_{1g}$, and $B_{2g}$.} In $B_{1g}$ symmetry, the spectral shapes close to the two-magnon maximum are universal. If Higgs fluctuations of the magnetization are included all three compounds can be described consistently \cite{Weidinger:2015}. However, since the spectral shape of La214 changes slightly with the photon energy additional contributions to the cross section may also exist including the triple resonance \cite{Chubukov:1995a,Chubukov:1995b,Morr:1997} or multi-magnon or chiral spin excitations \cite{Knoll:1990,Shastry:1990,Vernay:2007}. In spite of these open issues the $B_{1g}$ spectra of the cuprates can be considered understood by now including the numerical value of $J$ which agrees with the neutron results to within a few percent. Yet, the other symmetries which display high intensities as well are still far from being understood, in particular the role of chiral excitation remains elusive and cannot be treated with field-theoretical methods.

\begin{acknowledgments}
We acknowledge very fruitful discussions with P. Knoll, S. Weidinger, and W. Zwerger. Financial support for the work came from the DFG via the Transregional Collaborative Research Center TRR\,80  and the Bavarian Californian Technology Center BaCaTeC (project no. A5\,[2012-2]).
\end{acknowledgments}

\begin{appendix}
\label{sec:appendix}
\section{Surface treatment}
\label{Asec:surface}

\begin{figure}[h]
  \centering
  \vspace{-1mm}
  \includegraphics[width=0.85\columnwidth]{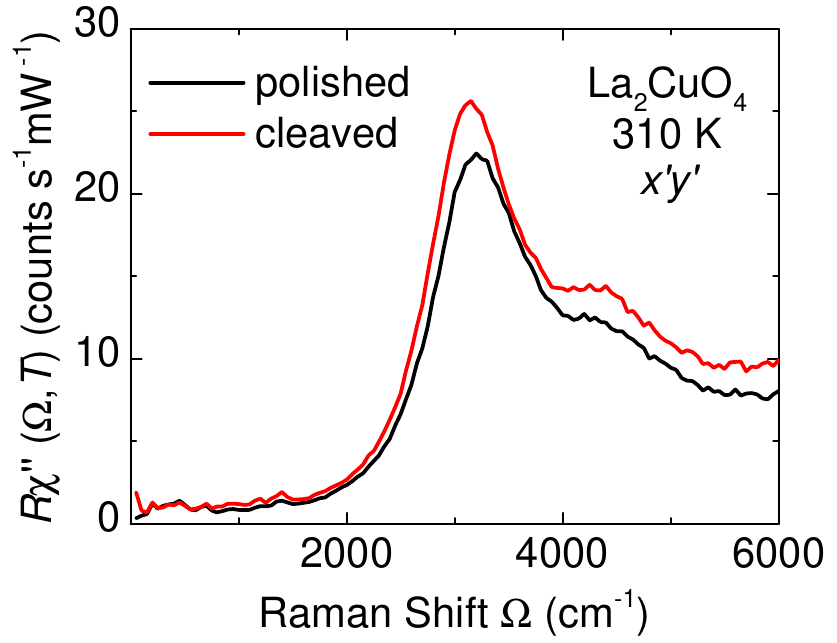}
  \vspace{0mm}
  \caption{Influence of the surface treatment in La214. There is a small intensity difference between the polished and the cleaved surface of the $B_{1g}$ spectra. All structures appear at the same position for both surfaces. The increase towards high energies for the cleaved sample has its origin in edge effects due to the small area available.
  }
  \vspace{-1mm}
  \label{sfig:cleave}
\end{figure}

The surface preparation is crucial only for La214 since natural surfaces can be used in the cases of Y123 and Bi2212:Y. For La214 polishing and cleaving are possible treatments \cite{Muschler:2010a}. However, polishing may result in damage and a severe influence on the spectra. On the other hand, a sufficiently large area is necessary for quantitative measurements and can only be obtained by polishing in the case of La214. For making sure that only intrinsic properties are observed we compared the results from the polished surface with those from a cleaved one which are usually reliable. However, La214 has no cleavage plane, and the flat areas necessary for maintaining the proper angle of incidence across the laser spot and avoiding wavelength dependent diffraction effects of the scattered photons.

When properly polished the differences between the two types of surfaces are small as shown in Fig.~\ref{sfig:cleave}. The intensity in the maximum is slightly higher for the cleaved surface. The positions of the main peak and the secondary maximum at 4600\,cm$^{-1}$ are nearly identical and the relative spectral weights as well. There is a strong increase of the intensity {towards high energies} for the cleaved surface. It is clear, however, that this increase is not intrinsic. Rather, the irregular shape of the surface and the relatively small flat part lead to various extrinsic contributions. For instance, both the incoming and the scattered photons are diffracted on edges close to the spot. In addition, since the surface close to the spot is not exactly flat, the angle of incidence varies across the spot. Therefore the light inside the bulk does not have a well defined polarization in cases other than $x$ or $y$. This complication completely hinders a symmetry analysis - apart from the other problems.

\section{Raw data}
\label{Asec:raw}

\begin{figure}[]
  \centering
  \vspace{0mm}
  \includegraphics[width=0.85\columnwidth]{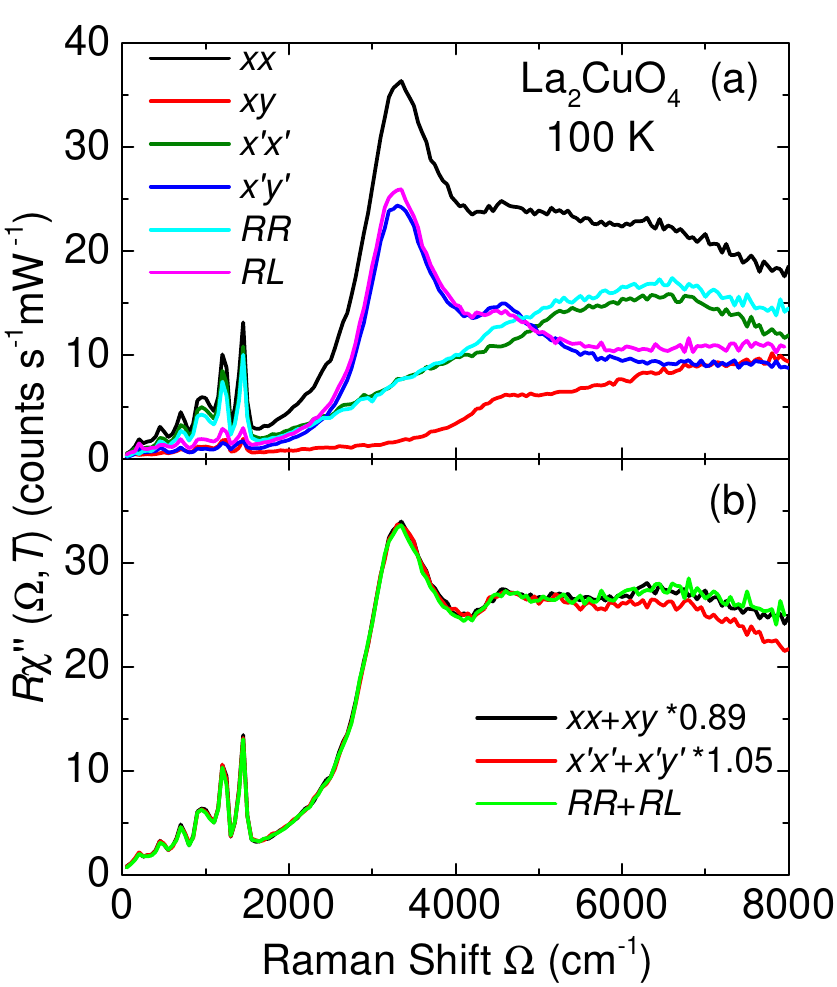}
  \vspace{0mm}
  \caption{Raman spectra of La$_2$CuO$_4$ at the six main polarization configurations (raw data after division by the Bose factor). (a) All spectra comprise two symmetries. (b) Each sum of two spectra with the same incident and orthogonal outgoing polarizations contains all four (in-plane) symmetries. Hence, the spectra shown should be equal. The discrepancies above 5000\,cm$^{-1}$ indicate extrinsic problems.
  }
  \label{sfig:raw}
\end{figure}

In Fig.~\ref{sfig:raw}\,(a) we show the raw data of La$_2$CuO$_4$ as measured at the six main polarization configurations $xx$, $xy$, $x^\prime,x^\prime$, $x^\prime,y^\prime$, $RR$, and $RL$ (in Porto notation). Each configuration contains two symmetries such as $A_{1g}+B_{1g}$ and $A_{2g}+B_{2g}$ for $xx$ and $xy$, respectively, hence the sum of two having the same incident and two orthogonal scattering polarizations comprises always the full set of in-plane symmetries. Therefore, one can check the consistency of the experiments as shown in Fig.~\ref{sfig:raw}\,(b). With small corrections, which result from insufficient adjustment of the absorbed laser power, the agreement in the energy range up to 5000\,cm$^{-1}$ is within the statistical error. Above 5000\,cm$^{-1}$ there are discrepancies between spectra with the incident polarization along $x^\prime$ and the other two sets. The agreement is better for Y123 and Bi2212, and the origin of these discrepancies is hard to pin down. Polarization dependent contributions from luminescence but also accumulating surface layers are possible explanations. In this range the derived spectra cannot be considered quantitative. In addition, polarization-independent contributions from luminescence are very like to occur in systems with defects in the crystal structure where charges can be trapped \cite{Bahrs:2005}.

\begin{figure}[h!]
  \centering
  \vspace{0mm}
  \includegraphics[width=0.85\columnwidth]{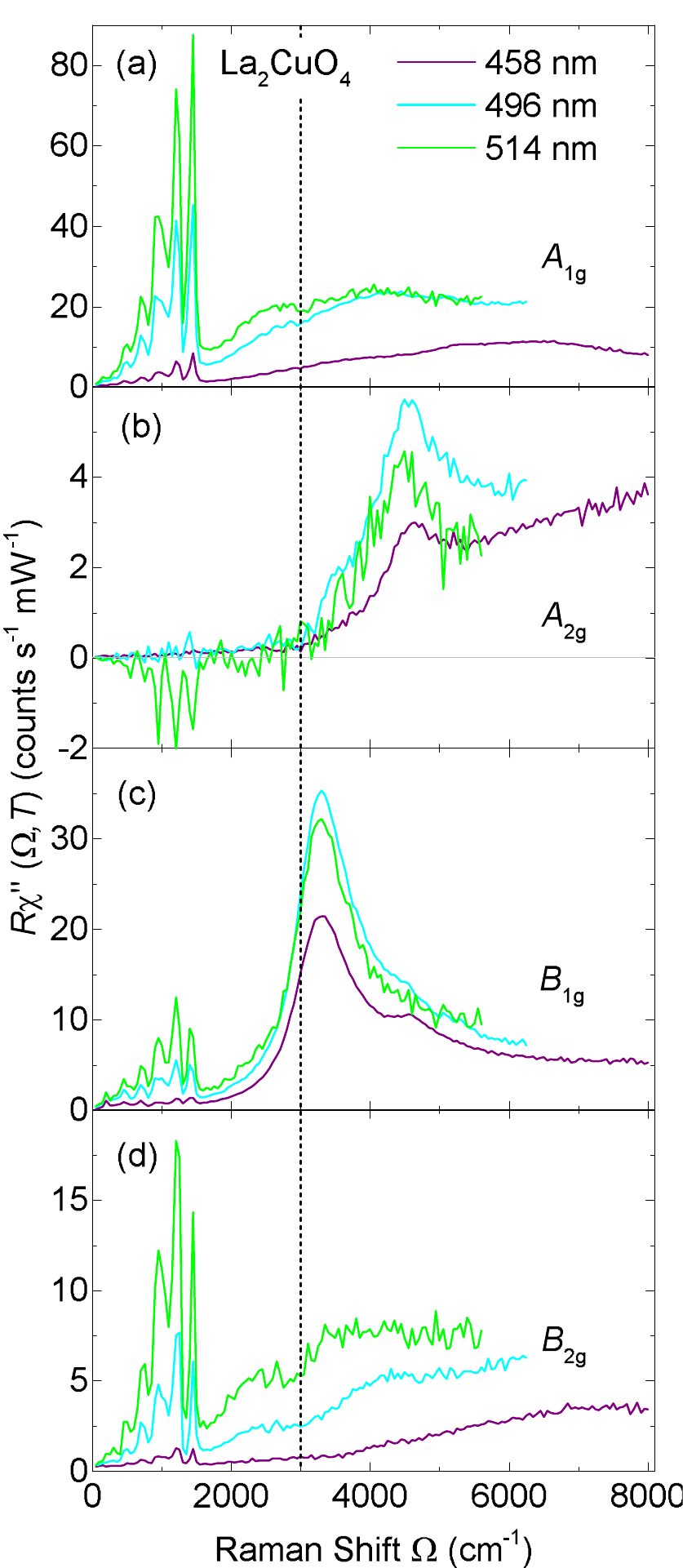}
  \vspace{0mm}
  \caption{{Symmetry-resolved Raman spectra of La$_2$CuO$_4$ at 100\,K and excitation energies as indicated. (b) and (c) The highest intensity of the $A_{2g}$ and $B_{1g}$ spectra is observed for $\hbar\omega_I = 2.5$\,eV (496\,nm). (a) and (d) In $A_{1g}$ and $B_{2g}$ symmetry the overall intensity increases monotonously towards low excitations energies, and the shape changes qualitatively. (b) The onset of the $A_{2g}$ intensity at 3000\,cm$^{-1}$ (dashed vertical line) is independent of the excitation. The peak energy varies slightly for reasons we do not know.}
  }
  \label{sfig:resonance-sym}
\end{figure}

\section{{Symmetry dependence of the resonances}}
\label{Asec:resonance-sym}

{For La214 we measured all pure symmetries for three different excitations lines at $\hbar\omega_I =  2.41$, 2.50, and 2.71\,eV (514.5, 496.5, and 457.9\,nm) as shown in Fig.~\ref{sfig:resonance-sym}. The first symmetry analysis was performed for Gd214 using the line at 488\,nm alone. \cite{Sulewski:1991} Except for details, partially coming from the lower temperatures here, but also from variations across the samples the results in Ref.~\onlinecite{Sulewski:1991} are close to our results measured with 496\,nm excitation.}

\begin{figure}[]
  \centering
  \vspace{0mm}
  \includegraphics[width=0.98\columnwidth]{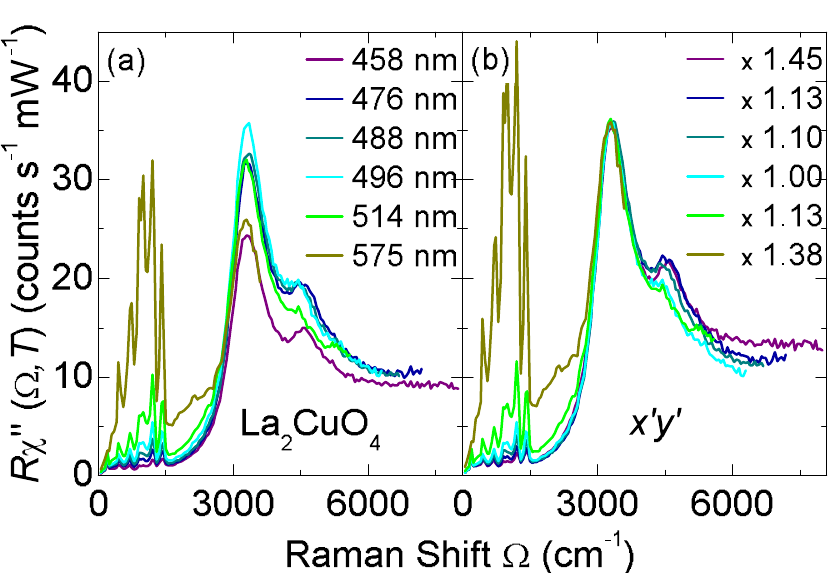}
  \vspace{0mm}
  \caption{{$A_{2g}+B_{1g}$ ($x^\prime y^\prime$) spectra at 100\,K for various excitation lines as indicated. (a) Raw data (including the calibration and Bose correction). The phonons below 1500\,cm$^{-1}$ resonate strongly towards the lower excitation energies. The two-magnon peak intensity is maximal for $\hbar\omega_I = 2.5$\,eV (496\,nm). The secondary maximum close to 4600\,cm$^{-1}$ becomes monotonously weaker between 2.71 and 2.41\,eV and is out of the range of the spectrometer for $\hbar\omega_I = 2.16$\,eV (575\,nm). (b) Normalized spectra. Close to the two-magnon excitation the shape is nearly universal in the range of photon energies studied here.}
  }
  \label{sfig:resonance-x'y'}
\end{figure}

{The line shapes in the range of the two-magnon scattering (2000-5000\,cm$^{-1}$) in $A_{2g}$ and $B_{1g}$ symmetry are only weakly dependent on $\hbar\omega_I$ [Fig.~\ref{sfig:resonance-sym} (b) and (c)] In either case the maximal intensity is found for the laser line at 496\,nm. The onset of the intensity in $A_{2g}$ symmetry is independent of the laser energy indicating an intrinsic inelastic process to set in above a gap of approximately 3000\,cm$^{-1}$ [Fig.~\ref{sfig:resonance-sym} (b)]. Since the variation with excitation energy is similar to that of the two-magnon peak in $B_{1g}$ symmetry the $A_{2g}$ response is likely to be excited via the same intermediate electronic states. The most probable explanation of the $A_{2g}$ response is a  chiral spin excitation. The variation with excitation energy makes it unlikely that the small peak on the high-energy side of the $B_{1g}$ peak [Fig.~\ref{sfig:resonance-sym} (c)] is related to the $A_{2g}$ response.}

{The shapes of the $A_{1g}$ and $B_{2g}$ spectra depend substantially on the incoming energy, and the overall intensities become monotonously stronger as opposed to those of the $A_{2g}$ and $B_{1g}$ spectra. The phonons are strongly resonant towards the green as observed earlier \cite{Sugai:1990}. Part of the variation in the range between 1500 and 3000\,cm$^{-1}$ originates most probably from resonant multi-phonon processes. Above 3000\,cm$^{-1}$ luminescence seems also to have an influence and is obviously not flat and symmetry independent.}

\section{{Resonance behavior of the $x^\prime y^\prime$ spectra}}
\label{Asec:resonance-x'y'}

{In Fig.~\ref{sfig:resonance-x'y'} we show the detailed resonance behavior of the spectra measured in $x^\prime y^\prime$ configuration corresponding to $A_{2g}+B_{1g}$ symmetry. In the main text in Fig.~\ref{fig:B1g_res} we show only spectra for two laser lines. Similar as in $A_{1g}$ symmetry the phonons resonate towards the green and become very strong for yellow excitation. Note that the individual lines cannot be resolved for the large steps between the measurement points. The two main observations being particularly clear in the normalized spectra [Fig.~\ref{sfig:resonance-x'y'}\,(b)] are (i) that the shape in the range around the two-magnon peak is only weakly varying and (ii) that the secondary maximum close to 4600\,cm$^{-1}$ fades away for green excitation. We could not measure the entire spectra with yellow light since the range of the spectrometer is limited.}

\end{appendix}

\end{document}